\documentclass[prd,aps,floats,twocolumn]{revtex4}

\usepackage[dvips]{graphicx}
\usepackage{amssymb}
\usepackage{amsmath}
\begin{document}

\title{Inflationary predictions for scalar and tensor fluctuations
  reconsidered}

\author{Latham A. Boyle$^1$, Paul J. Steinhardt$^1$, and Neil
  Turok$^2$} 

\affiliation{$^1$Department of Physics, Princeton University,
  Princeton, New Jersey 08544 \\
  $^2$Department of Applied Mathematics and Theoretical Physics,
  Centre for Mathematical Sciences, University of Cambridge,
  Wilberforce Road, Cambridge CB3 OWA, United Kingdom}

\date{July 2005}

\begin{abstract}
  We reconsider the predictions of inflation for the spectral index of
  scalar (energy density) fluctuations ($n_s$) and the tensor/scalar
  ratio ($r$) using a discrete, model-independent measure of the
  degree of fine-tuning required to obtain a given combination of
  ($n_s$, $r$).  We find that, except for cases with numerous
  unnecessary degrees of fine-tuning, $n_s$ is less than $0.98$,
  measurably different from exact Harrison-Zel'dovich.  Furthermore,
  if $n_s \gtrsim 0.95$, in accord with current measurements, the
  tensor/scalar ratio satisfies $r\gtrsim 10^{-2}$, a range that
  should be detectable in proposed cosmic microwave background
  polarization experiments and direct gravitational wave searches.
\end{abstract}
\maketitle 

Inflation predicts nearly scale-invariant spectra of primordial scalar
(energy density) and tensor (gravitational wave) perturbations.  What
has been less clear is the precise prediction for the scalar spectral
index $n_s$ and the tensor/scalar ratio $r$. In particular, is $n_s$
likely to be distinguishable from pure Harrison-Zel'dovich ($n_s=1$)?
And is $r$ likely to be large enough for the tensor perturbations to
be detected ($r \gtrsim 10^{-2}$)?  One approach for addressing these
questions is anecdotal experience based on explicitly constructing
inflaton potentials $V(\phi)$ with different combinations of ($n_s$,
$r$).  A more recent approach is to use the inflationary flow
equations to compute $n_s$ and $r$ for random choices of the
``Hubble'' slow-roll parameters \cite{flow_eqs}, and plot the results
as a dot-plot in the ($n_s$, $r$) plane.  One problem with these
methods is that the sampling does not incorporate a weight based on
physical plausibility, so it does not provide a well-motivated measure
of the relative likelihood across the ($n_s$, $r$)-plane. It is as if
all inflaton potentials are created equal.  Another problem in many of
these studies is that only some of the minimal requirements for a
successful inflaton potential are considered.  It is simply assumed
that the rest can be satisfied without reducing the attractiveness of
the model.  Yet, this assumption is often invalid in practice.

In this Letter, we attempt to rectify this situation by considering
the complete set of inflationary conditions and introducing a discrete
counting scheme for assessing the degrees of fine-tuning required to
obtain a given combination of $n_s$ and $r$.  (An alternative approach
is based on Bayesian model selection \cite{trotta}.)  We find (see
Fig.~1) that models with blue or slightly red tilts ($n_s > 0.98$)
require significantly more degrees of fine-tuning than models with
$n_s < 0.98$.  This concurs with the intuitive impression obtained by
trying to construct potentials by hand.  More importantly, the
procedure reveals valuable additional information: (1) a significant
gap exists between the inflationary prediction for $n_s$ and pure
Harrison-Zel'dovich ($n_s=1$), a difference that near-future
measurements should be able to resolve; and (2) if $n_s \gtrsim 0.95$,
as current measurements suggest, $r$ exceeds $10^{-2}$, so tensor
fluctuations should be observable in proposed cosmic microwave
background (CMB) polarization and gravitational wave interferometer
experiments.  (Interestingly, independent approaches based on physical
field-theoretic arguments have reached similar conclusions
\cite{deVegaSanchez}.)

As we have emphasized, it is important to consider all the conditions
necessary for inflation when assessing the degrees of fine-tuning,
namely:
\begin{enumerate}
\item as $\phi$ evolves over some range $\Delta \phi$, the universe
  undergoes at least $N>60$ $e$ folds of inflation in order to become
  homogeneous, isotropic, spatially flat, and monopole-free;
\item after the field evolves past this range, inflation must halt and
  the universe must reheat without spoiling the large-scale
  homogeneity and isotropy;
\item the energy density (scalar) perturbations, which we assume are
  generated by the quantum fluctuations of the inflaton field, must
  have amplitude $\sim 10^{-5}$ on scales that left the horizon
  $\approx 60$ $e$ folds before the end of inflation, to agree with
  observations \cite{reheat};
\item after inflation, the field must evolve smoothly ({\it i.e.},
  without generating unacceptable inhomogeneities) to an analytic
  minimum with $V \approx 0$;
\item if the minimum is metastable, then it must be long-lived and $V$
  must be bounded below.
\end{enumerate}
The analyticity condition is to avoid physically questionable terms of
the form $|\phi|$ or $\phi^{\alpha}$ where $\alpha$ is non-integer.
Many analyses consider only the first three conditions, but we find
that the fourth condition, which is equally essential, imposes a
non-linear constraint on $V$ that can significantly affect the degree
of fine-tuning required to obtain a given ($n_s$, $r$).  (We have
stated the conditions above as if the inflaton potential is a function
of a single field $\phi$; the generalization to multiple fields is
straightforward.)

To quantify the degree of fine-tuning, we count the number of
unnecessary features introduced during the last 60 $e$ folds of
inflation to achieve a given ($n_s$, $r$).  To pose the conditions in
a physically motivated and model-independent way, we use the standard
slow-roll parameters:

\begin{eqnarray}
  \label{eps_eta}
  \epsilon & \equiv & (3/2)(1+w) \approx 
  (1/2) d \, {\rm ln} \, V/d N  \\
  \eta & \equiv &  (1/2) d \, {\rm ln} \, (V')^2/d N,
\end{eqnarray}
where $N$ is the number of $e$ folds remaining before inflation ends
and a prime indicates $d/d\phi$ for an inflaton field $\phi$
canonically normalized in Einstein frame.  The parameters $\epsilon$
and $\eta$ have a physical interpretation: they represent respectively
the fractional rate of change of the Hubble parameter ($\propto
V^{\frac{1}{2}}$) and the force on the scalar field ($\propto V'$) per
inflationary $e$ fold.  In all inflationary models, $\epsilon$ and
$\eta$ must increase from small values ($\lesssim 1/60$) when $N
\approx 60$ to values of order unity at the end of inflation ($N=0$).

For minimally tuned models, the simplest being a monomial potential
$V= \alpha \phi^n$ with integer $n$ and a single adjustable
coefficient $\alpha$, both $\epsilon(N)$ and $\eta(N)$, as well as all
of their derivatives ({\it e.g.}, $d^m \eta/dN^{m}$) are monotonic and
have no zeroes during the last 60 $e$ folds.  The range of ($n_s$,
$r$) associated with these models lies in the shaded region marked
``0" in Fig.~1, which has $n_s < 0.98$ and $r > 10^{-2}$.  To move
further outside this range requires that more zeroes of $\eta$ and its
derivatives occur in the last 60 $e$ folds.  The zeroes are
independent in the sense that they can be added one at a time by
successively adjusting parameters, as shown in Fig.~1.

Our key point is this: As exemplified by the minimally tuned models,
no zeroes whatsoever are required during the last 60 $e$ folds to
satisfy the five inflationary conditions.  Hence, each zero added to
the last 60 $e$ folds can be properly construed as representing an
extra degree of fine-tuning beyond what is necessary -- an extra
acceleration, jerk or higher order-shift in the equation of state (for
$\epsilon$) or the force (for $\eta$) artificially introduced at
nearly the exact moment when the modes currently observed in the
cosmic microwave background are exiting the horizon during inflation.

More specifically, the number of zeroes is a conservative (lower
bound) measure of how many derivatives of $\epsilon(N)$ and $\eta(N)$
must be finely adjusted to achieve a given $(n_s, r)$.  This can be
seen by constructing the Taylor expansion about $N_0 \approx 60$ and
comparing the higher-order terms to the lower order ones at, say,
$N=10$.  As the number of zeroes increases, more higher-order terms
contribute non-negligibly before inflation ends, revealing the
delicate toggling of the equation of state and the force $V'$ during
the last 60 $e$ folds.  Proceeding deeper into the gray region in
Fig.~1 (many tunings), the terms in the Taylor series grow until the
series is no longer absolutely convergent.

Therefore, we propose to quantify the fine-tuning by introducing the
integers $Z_{\epsilon,\eta}$ which measure the number of zeroes that
$\epsilon$ and $\eta$ and their derivatives undergo within the last 60
$e$ folds of inflation.  Fig.~1 is based on zeroes of $\eta$; a
similar result occurs for $\epsilon$.  We find that these metrics are
robust methods for dividing models into those that are simple (few
zeroes) and those that are highly tuned (many zeroes).  (N.B.  The
point of our metric is not to rank a model with $Z_{\eta}=20$ over a
model with $Z_{\eta}=1000$; the significance of a difference in
$Z_{\eta}$ when $Z_{\eta}$ is large for both models is unclear.
Rather, $Z_{\eta}$ is designed to show that both models are vastly
more finely adjusted than models with $Z_{\eta}=0$ or 1.)

\begin{figure}
  \begin{center}
    \includegraphics[width=3.1in]{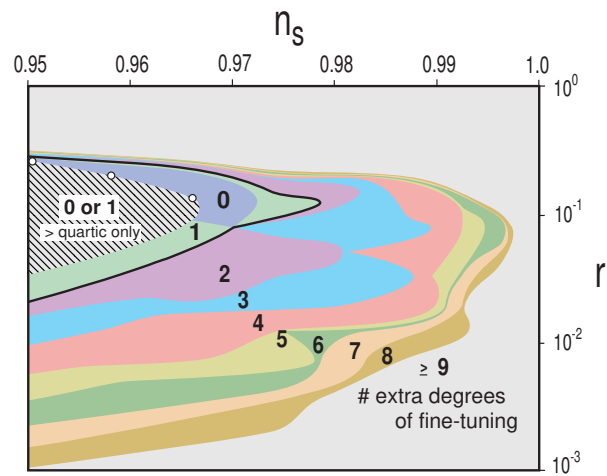}
  \end{center}
  \caption{Predictions for tensor/scalar ratio $r$ versus spectral tilt 
    $n_{s}$ for minimal tuning ($Z_{\eta}=0$) and for different
    degrees of extra fine-tuning ($Z_{\eta} \ge 1$).  The small white
    circles correspond to monomial potentials (from right to left:
    quadratic, cubic, quartic). The thick curve enclosing all models
    with zero or one extra degree of fine-tuning has $n_s < 0.98$ and
    $r> 10^{-2}$; hatched portion has $Z_{\eta} =0$ or~1 but is only
    accessible for polynomials of degree greater than four ($Z_{order}
    \ge 1$). Nine or more extra degrees of fine-tuning are required to
    obtain $n_s$ close to 1 or small $r$ (gray).}
  \label{rVtilt_plot}
\end{figure}

Fig.~1 summarizes our analysis for quartic polynomial potentials
$V(\phi)$ that satisfy the five inflationary conditions.  The simplest
models, the monomial potentials, are represented by a sequence of
discrete white circles.  Next, we consider more general polynomials
combining terms of different order.  The cases in which the
coefficients all have the same sign lie on the boundary of the shaded
region marked ``0", along the curve connecting the white circles.  All
of these models are minimally tuned ($Z_{\eta}=0$) and have $n_s <
0.97$ and $r> 10\%$.

A special case occurs among models with only one degree of fine-tuning
$Z_{\eta}=1$: namely, models tuned so that the 60 $e$-fold mark lies
very near a maximum of the potential.  Simple examples include the
Mexican hat potential, $V(\phi)= V_0 - \frac{1}{2} m^2 \phi^2 +\lambda
\phi^4$ and the axion potential, $V(\phi) = V_0 (\,{\rm 1+
  cos}(\phi/f))$.  If $\phi$ at the 60 $e$-fold mark lies close to the
maximum, then $ \eta$ has a zero since the force must have a maximum
in these potentials.  Although this kind of zero is unnecessary for
inflation, it can occur naturally if the action is invariant under
certain symmetries, as illustrated by the two examples above. Hence,
we include this region within our thick black curve in Fig.~1.  As the
parameters of $V$ are further adjusted so that $\phi$ lies very close
to the maximum at the 60 $e$-fold mark, the allowed range in the
($n_s$, $r$) plane expands to fill out the shaded region marked ``1."

Everywhere else in the ($n_s$, $r$) plane is reached by adding
sequentially more zeroes of $\eta$ and its derivatives within the last
60 $e$ folds.  Increasing the zeroes introduces one or more special
features in $V$ (extrema, inflection points, $\ldots$), progressively
flattens the potential in the vicinity of the feature, and finely
tunes $\phi$ at the 60 $e$-fold mark to lie closer and closer to it.
Unlike the first ($Z_{\eta}=1$) case of tuning discussed above, there
is no symmetry principle that dictates any of these additional
tunings.  Yet, as Fig.~1 shows, many such tunings are necessary to
reach low values of $r$ or high values of $n_s$.
  
Although Fig.~1 is based on quartic (renormalizable) polynomial
potentials, a similar plot can be constructed for polynomials of
arbitrary order.  For polynomials of any order, there is always a wide
range of parameters for which $Z_{\eta}=0$~or~1, $n_s \lesssim 0.98$
and $r \gtrsim 10^{-2}$. With higher-order polynomials, it is possible
to insert more bumps and jerks into the final 60 $e$ folds, even
though this is not required for inflation.  Introducing them for the
purpose of enabling anomalous values of $n_s$ and $r$ should be
included in assessing the degrees of fine-tuning.  Just as the zeroes
are independent and can be added one by one, the space of polynomial
functions can be extended order by order. Hence, we suggest amending
the degrees of tuning to be $Z_{\eta} +Z_{order}$, where $Z_{order}$
is the difference between the actual polynomial order and four.  One
regime that now becomes accessible with $Z_{\eta}=0$ or~1 is the
hatched region in Fig. 1; $n_s \gtrsim 0.98$ and/or $r \lesssim
10^{-2}$ (for $n_s \gtrsim 0.95$) still require many degrees of
fine-tuning.

Models with more than one field can be treated in a similar way
provided the path(s) describing the last 60 $e$ folds of inflation and
the passage to the potential minimum can be described by the classical
equations of motion for the fields.  We shall call these
``deterministic."  The above analysis may simply be applied to each
path individually.  As before, each path with many zeroes is related
to paths with $Z_{\eta}=0$~or~1 by a continuous fine-tuning of
parameters.  In some special models (like the hybrid model in
\cite{hybrid}), the path is non-deterministic.  Instead, the classical
evolution reaches a critical point in the potential where quantum
diffusion is needed to reach the end of inflation, as in the case
$V=V_0 + \frac{1}{2} m_{\phi}^2 \phi^2 - \frac{1}{2} m_{\psi}^2 \psi^2
+ \frac{1}{2} \gamma \phi^2 \psi^2 + \ldots$ (which has a critical
point at $\gamma \phi_c^2 = m_{\psi}^2$ and $\psi=0$).  There is
effectively a discontinuous jump in $\epsilon$ and $\eta$ at the
critical point, and there is not a unique procedure to relate these
cases to models with $Z_{\eta}=0$~or~1.  As a result, counting zeroes
may not be an appropriate way to judge them.  We note that these cases
include examples with $n_s \gtrsim 0.98$ and $r \lesssim 10^{-2}$;
however, compared to the $Z_{\eta}=0$ models in Fig.~1, one must add
at least one extra field, an exponentially large mass hierarchy
between $m_{\phi}$ and $m_{\psi}$, and, for some ($n_s$, $r$), another
hierarchy between the three dimensionless quartic couplings and/or one
or more higher-order couplings \cite{baumann}.

Our conclusion that gravitational waves should be detectable runs
contrary to some claims in the literature.  A common but flawed
argument has been that the amplitude of the tensor power spectrum
is highly uncertain because it is proportional to the fourth power of
the inflationary energy scale $M_*$, whose value is poorly determined.
In actuality, although gravitational waves will allow us to determine
$M_*$, their detectability only depends on the tensor/scalar ratio,
$r$, which does not depend on $M_*$ at all \cite{myth}.
($r \approx 16 \epsilon$, so it only depends on the equation of state
during inflation.)

A second argument by Lyth \cite{lyth_bound} and others makes the claim
that $r$ must be small in any theory which includes quantum gravity
effects. They point out that the slow-roll equations imply the
relation $\Delta \phi = (m_{Pl}/8 \sqrt{\pi}) \int r^{1/2} dN$, where
$m_{Pl} = 1.2 \times 10^{19}$~GeV is the Planck mass. From this
relation, if $r\gtrsim .05$, then $\Delta \phi/ m_{Pl}$ should exceed
unity over the final 60 $e$ folds of inflation. Lyth argues that once
gravitational effects are included: (a) the effective potential $V$
can only be reliably calculated over a domain $\Delta \phi<m_{Pl}$;
and (b) inflation is likely to occur only over a range $\Delta
\phi<m_{Pl}$.  These two claims are disputed in \cite{deVegaSanchez},
and several inflationary models \cite{Lyth_counter_examples} provide
explicit counter-examples.  These models have $\Delta \phi/ m_{Pl} >
1$ and gravitational corrections are under control. They give values
of $n_s$ and $r$ consistent with Fig.~1 and with current data.

Our analysis shows that forcing $\Delta \phi/m_{Pl}$ to be less than
unity and maintaining a small number of zeroes requires $n_s \lesssim
0.95$, past the left boundary of Fig.~1 and outside the range favored
by current data.  To obtain $r < 10^{-3}$ and a small number of zeroes
requires much smaller $n_s$. The only alternatives for obtaining small
$r$ are to introduce many zeroes or to turn to non-deterministic
models.  The latter, as we have noted, typically require extra fields
and tunings compared to deterministic models with $Z_{\eta}=0$ or~1.

The primordial spectrum of gravitational waves in Fig.~1 is related to
the observable spectrum today ($\tau_0$) by a ``transfer function''
$T_{t}(k,\tau_{0})$ \cite{Boyle05}:
\begin{equation}
  \Omega_{gw}(k,\tau_{0}^{})\equiv\frac{1}{\rho_{cr}^{}}
  \frac{d\rho_{gw}^{}}{d\,{\rm ln}\,k}=\frac{1}{12}\frac{k^{2}}
  {a_{0}^{2}H_{0}^{2}}T_{t}(k,\tau_{0})\Delta_{t}^{2}(k),
\end{equation}
where $\Delta_{t}^{2}(k)$ is the primordial tensor power spectrum,
$\rho_{gw}$ is the gravitational-wave energy density, $\rho_{cr}$ is
the critical density, and $\Omega_{gw}(k, \tau_0)$ is the ratio of the
gravitational wave energy density in a log-interval about $k$ to the
critical density.  We have used an analytic expression for the
transfer function derived in \cite{Boyle05} which improves on the
accuracy of previous calculations \cite{Turner_gwave} (but see also
\cite{kamionkowski, bashinsky}).  The transfer function includes the
redshift-suppression after horizon re-entry; the imprint of horizon
re-entry itself; the possibility of dark energy with equation-of-state
$w(z)$; the damping due to free-streaming relativistic particles ({\it
  e.g.}, neutrinos) in the early universe \cite{Weinberg}; and several
early-universe effects that were not considered in previous
treatments.

\begin{figure}
  \begin{center}
    \includegraphics[width=3.4in]{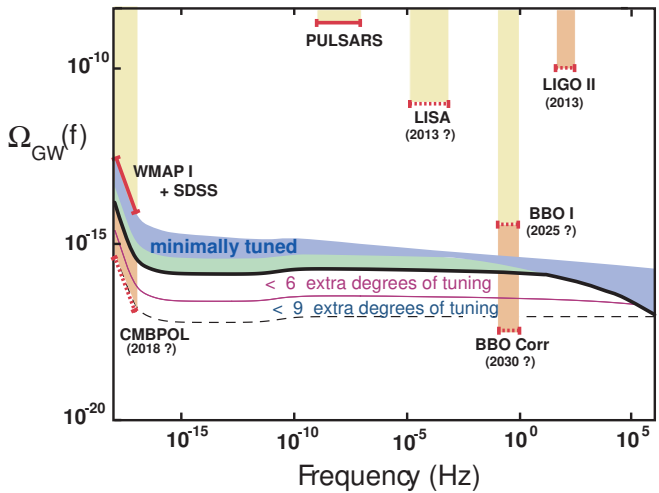}
  \end{center}
  \caption{Inflationary predictions of $\Omega_{gw}(f)$ vs.\ $f$
    with present (solid bars) and future (dashed bars) observational
    limits.  The solid blue and green bands represent the predicted
    range for models with minimal tuning, or one extra degree of
    tuning, respectively, that satisfy the current observational
    bounds.  The thick curve represents the lower-bound for
    $\Omega_{gw}$ from among the region enclosed by the black curve in
    Fig.~1.  The purple curve is the lower bound for models with
    $Z_{\eta}<6$.  The dashed curve ($r=10^{-3}$) is the lowest
    prediction among all models shown in Fig.~1.}
  \label{energy_plot}
\end{figure}

Fig.~2 shows the inflationary predictions for $\Omega_{gw}$ as a
function of frequency $f$ compared to present and future observations,
assuming current limits, from \cite{Seljak}, on non-inflationary
parameters.  (For a related figure with the observations shown in more
detail, see Fig.~2 in \cite{kamionkowski}.)  The thick solid curve in
Fig.~2 represents the lower bound among all models with minimal tuning
($Z_{\eta}=0$) or one extra degree of fine-tuning ($Z_{\eta}=1$).  The
thick dotted curve is the lower bound predicted for the entire range
of models in Fig.~1; for quartic potentials, at least nine extra
degrees of fine-tuning are required to go below it.  Also notice the
kinks near $f\sim 10^{-11}$~Hz, caused by the onset of neutrino
free-streaming, as reflected in the tensor transfer function.  Most
importantly, the entire range of models discussed here should be
accessible to future CMB polarization experiments
\cite{Planck,Verde,report} and space-based gravitational wave
detectors, like the Big Bang Observer (BBO) \cite{BBO}.

Hence, we find that, contrary to some suggestions in the literature,
all inflationary models are not created equal.  The goals of inflation
do not require going beyond models with minimal or near-minimal tuning
($Z_{\eta} \le 1$).  (For skeptical readers who may demur from this
conclusion, we pose a challenge: construct a deterministic, complete
inflationary model forced by fundamental physics into a parameter
region with $Z_{\eta} \gg 1$.)  Furthermore, the minimal and
near-minimal models are the most powerfully predictive in the sense
that they require the fewest tunings and make the highest number of
successful predictions.  Based on this analysis, both a red tilt with
$n_s < 0.98$ and cosmic gravitational waves with $r \gtrsim 10^{-2}$
are expected and should be detected if inflation is right.  A similar
analysis should be applied to cyclic models \cite{cyclic}, which must
satisfy some conditions analogous to the inflationary conditions (and
some not), to determine if the same range of $n_s$ is favored.

We thank L.~Page and D.~Spergel for discussions that inspired this
project, and D.~Baumann for many insightful comments.  This work was
supported in part by US Department of Energy grant DE-FG02-91ER40671.


\begin{thebibliography}{99}
\bibitem{reheat} Although the text states $N_0\approx 60$, we compute
  the precise value of $N_0$ model by model in our analysis.  Also, we
  assume instantaneous reheating, but it is straightforward to show
  that assuming a low reheat temperature (or preheating) does not
  change our conclusions.
\end{thebibliography}
\end{document}